\documentclass[prl,twocolumn,superscriptaddress,showpacs,preprintnumbers,amsmath,amssymb,floatfix]{revtex4-1}
\usepackage{hyperref}
\usepackage{epsfig}

\newcommand{\be}{\begin{equation}}
\newcommand{\ee}{\end{equation}}

\newcommand{\md}[1]{\left|#1\right|}
\newcommand{\Ima}{\mathrm{Im}}

\begin{document}

\title{Noncontact dissipation reveals critical central peak  in  SrTiO$_3$}
\author{M. Kisiel}
\affiliation{Department of Physics, University of Basel, Klingelbergstrasse 82, 4056 Basel, Switzerland}
\author{F. Pellegrini}
\affiliation{SISSA, Via Bonomea 265, I-34136 Trieste, Italy}
\affiliation{CNR-IOM Democritos National Simulation Center, Via Bonomea 265, I-34136 Trieste, Italy}
\author{G.E. Santoro}
\affiliation{SISSA, Via Bonomea 265, I-34136 Trieste, Italy}
\affiliation{CNR-IOM Democritos National Simulation Center, Via Bonomea 265, I-34136 Trieste, Italy}
\affiliation{International Centre for Theoretical Physics (ICTP), P.O.Box 586, I-34151 Trieste, Italy}
\author{M. Samadashvili}
\affiliation{Department of Physics, University of Basel, Klingelbergstrasse 82, 4056 Basel, Switzerland}
\author{R. Pawlak}
\affiliation{Department of Physics, University of Basel, Klingelbergstrasse 82, 4056 Basel, Switzerland}
\author{A. Benassi}
\affiliation{ Empa, Swiss Federal Laboratories for Materials Science and Technology, \"Uberlandstrasse 129, 8600 D\"ubendorf, Switzerland \\}
\affiliation{Institute for Materials Science and Max Bergmann Center of Biomaterials, TU Dresden, 01062 Dresden, Germany\\}
\author{U. Gysin}
\affiliation{Department of Physics, University of Basel, Klingelbergstrasse 82, 4056 Basel, Switzerland}
\author{R. Buzio}
\affiliation{CNR-SPIN Institute for Superconductivity, Innovative Materials and Devices, C.so Perrone 24, 16152 Genova, Italy}
\author{A. Gerbi}
\affiliation{CNR-SPIN Institute for Superconductivity, Innovative Materials and Devices, C.so Perrone 24, 16152 Genova, Italy}
\author{E. Meyer}
\affiliation{Department of Physics, University of Basel, Klingelbergstrasse 82, 4056 Basel, Switzerland}
\author{E. Tosatti}
\affiliation{SISSA, Via Bonomea 265, I-34136 Trieste, Italy}
\affiliation{CNR-IOM Democritos National Simulation Center, Via Bonomea 265, I-34136 Trieste, Italy}
\affiliation{International Centre for Theoretical Physics (ICTP), P.O.Box 586, I-34151 Trieste, Italy}

\date{\today}

\begin{abstract}
The critical fluctuations at second order structural transitions in a bulk crystal may affect the dissipation of mechanical probes 
even if completely external to the crystal surface. Here we show that noncontact force microscope dissipation bears clear evidence 
of the antiferrodistortive phase transition of SrTiO$_3$, known for a long time to exhibit a unique, extremely narrow neutron scattering 
``central peak''. The noncontact geometry suggests a central peak linear response coupling connected with strain.  
The detailed temperature dependence reveals for the first time the intrinsic
central peak width of order 80 kHz, two orders of magnitude below the established neutron upper bound.  
\end{abstract}
\maketitle

Second order structural phase transitions leave a clear mark in all thermodynamical, 
mechanical, equilibrium and non-equilibrium properties of bulk crystals. It was proposed some time ago  
that the critical fluctuations should also leave a footprint in the frictional dissipation of external mechanical probes  
such as an atomic force microscope (AFM) when temperature crosses the phase transition in the underlying bulk~\cite{Benassi11}.
The recent successful detection of a superconducting transition in the linear response mechanical dissipation  
of a noncontact, pendulum-type AFM tip hovering more than one nm above the sample surface ~\cite{Kisiel2011} 
suggests that continuous structural transitions might also be detectable in this manner. 
Here we present a first realization of this idea, with direct application to a most classic example, the
antiferrodistortive transition of SrTiO$_3$ just above 100 K.  At this phase transition the high temperature  
ideal cubic perovskite crystal structure becomes unstable against a zone-boundary phonon-like displacement 
of the ions, leading to a cell doubling and a tetragonal I4/mcm symmetry at lower temperatures. This exquisitely 
second order ``displacive'' transition historically provided a clean realization of nonclassical critical exponents~\cite{Muller71}. 
A very intriguing feature of this system, originally uncovered by neutron scattering, and later confirmed by 
other techniques, is the so-called ``central peak''~\cite{Riste71,Shapiro72, Reiter80, Shirane93}.
Very close to the critical transition temperature $T_c$, inelastic neutron spectra showed, besides ordinary critical 
fluctuations -- which proliferate and soften but never reach zero frequencies -- a  strikingly narrow peak 
(less than the 6 MHz width resolution) centered at zero frequency, whence the name. 
The central peak (CP) intensity appeared to obey the static critical exponents of the transition, but 
despite considerable efforts the actual nature and  width of the central peak were not uncontroversially 
established~\cite{Cowley96}.

Here we show that noncontact pendulum AFM dissipation, measured far from actual contact with the surface, reveals 
for the first time a structural phase transition, and it does so by revealing the CP of SrTiO$_3$. A linear response analysis 
shows that the CP-related mechanical loss peak %may be 
is
as narrow as 80 kHz, a frequency orders of magnitude 
below the neutron established upper bound.  Moreover, even if it cannot strictly determine %its
the intimate nature of the CP, 
the mechanical coupling suggests a connection with critical  fluctuations of strain, which are known to be 
associated with those of the main antiferrodistortive order parameter~\cite{Hochli80}.

The probe consisted of a very soft, highly doped silicon cantilever (ATEC-Cont from Nanosensors) with spring constant $k=0.1$ N/m, 
suspended perpendicularly to the surface with an accuracy of 1$^{\circ}$ and operated in the so-called pendulum 
geometry %at which 
where the tip vibration 
describes an arc
parallel to the sample surface.  
The peculiarity of this technique is to detect phenomena, in this case phase transitions, 
which happen in bulk, by means of a non-invasive, ultra sensitive and local surface probe, as opposed to  traditional probes 
such as neutrons and X-rays which invade the bulk in a much more global fashion. Moreover, the pendulum AFM is a kilohertz probe, 
sensitive to phenomena and to fluctuations that may take place on a much slower time scale than that accessible
with neutrons or X-rays. 
The oscillation amplitude $A$ of the tip was kept constant to approximately 5 nm using a phase-locked loop feedback 
circuit. The cantilever was annealed in UHV up to 700$^{\circ}$C for 12 h, which results in removal of water layers 
and other contaminants from both the cantilever and the tip. After annealing the cantilever quality factor, frequency 
and internal damping were equal to $Q=7\cdot 10^{5}$, $f_0=11$ kHz and $W_0=2\cdot 10^{-12}$ kg/s, respectively. 
%This long-term 
The annealing is also known to reduce all localized charges on the probing silicon tip \citep{Kisiel2011},
which is neutral, %meaning 
since the tip-sample contact potential difference was compensated (V=V$_{CPD}$) during the experiment.

\begin{figure}[!ht]\centering
\includegraphics[width=0.45\textwidth]{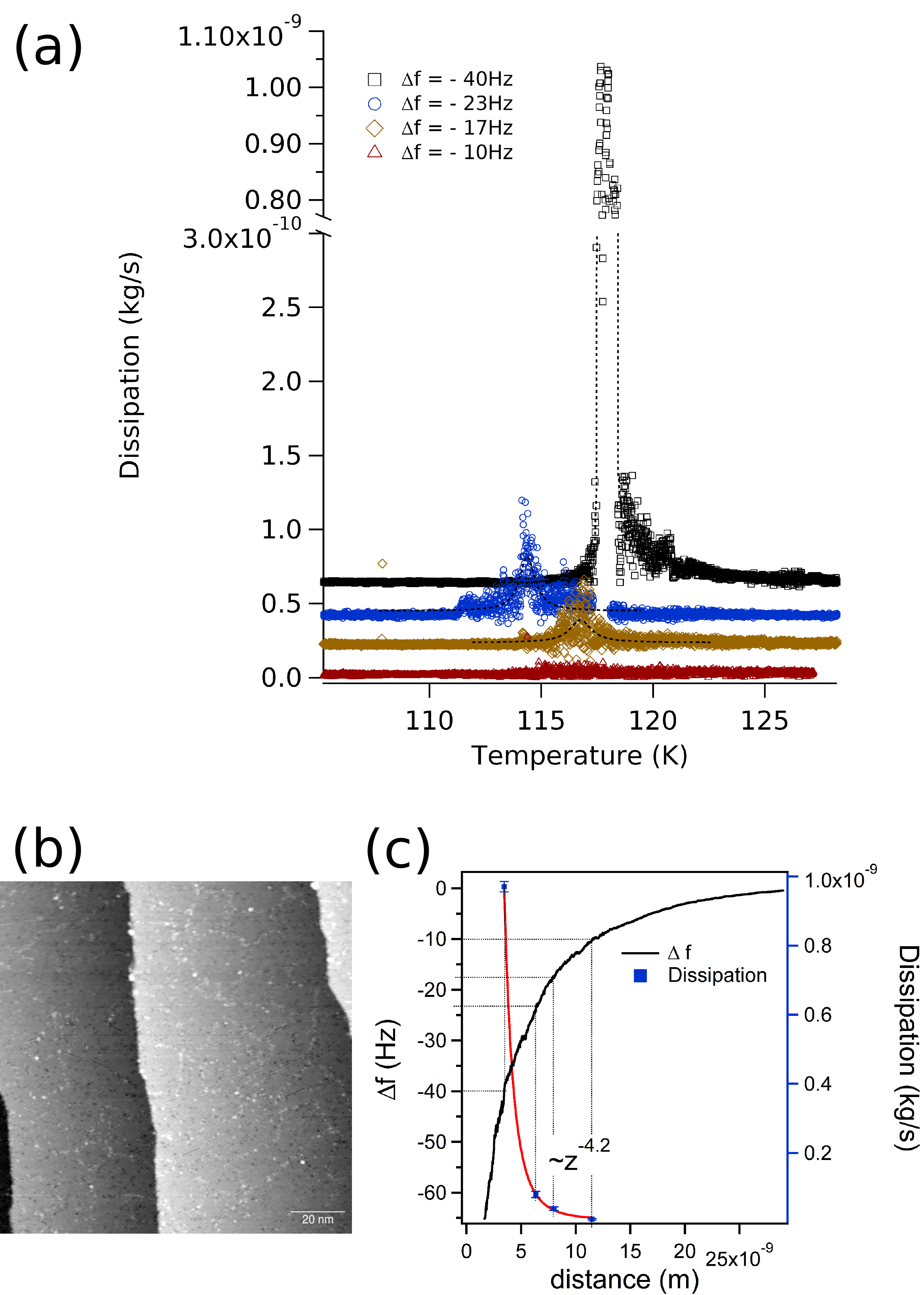}
\caption{\label{Exp_diss} 
(a) - Experimental AFM dissipation $W$ as a function of temperature. Raw data, taken at different
surface spots and different tip sample distances $z$. The sharp peak corresponds to the critical temperature of 
SrTiO$_3$ in the bulk region under the tip. 
(b) - Low temperature (T=5K) STM image of SrTiO$_3$(100) surface. The image is obtained at constant 
current I=10pA and bias voltage U=1V. More details in Supplementary)
(c) - The distance dependence of dissipation $W$, taken as the maximum of the peak shown in Fig.1(a), 
at four different spots on the sample. A fit to the experimental data $W \propto z^{-p}$ is shown in red, with $p \sim 4.2$. This exponent is close to the value $p=4$ 
expected for phononic dissipation, as appropriate for coupling to acoustical surface fluctuations 
of an insulating bulk material.}
\end{figure}

Fig.~\ref{Exp_diss} (a) shows the power $W(T)$ dissipated by the pendulum AFM as a function of temperature
at different spots on the SrTiO$_3$ surface and at different tip-sample distances, as measured by the
shift of the resonance frequency $\Delta f$.  
The dissipation is inferred from the standard expression \cite{gotsman99} 
$W=W_0\left(A_\mathrm{exc}(z)/A_{\mathrm{exc},0}-f(z)/f_0\right)$ in terms of the measured 
distance-dependent excitation amplitude $A_\mathrm{exc}(z)$ and resonance frequency $f(z)$ (where
$f(z)=f_0+\Delta f$) of the cantilever, the suffix $0$ referring to the free cantilever.
Since the Young modulus of the silicon cantilever is temperature dependent also the frequency of the free cantilever changes as a function 
of temperature - $\Delta f_0(T)$ \cite{gysin04}. %In the proximity of the sample, for 
In a temperature dependent experiment the total change of the frequency is $\Delta f(T,z)=\Delta f_0(T)+\Delta f(z)$ 
where $\Delta f(z)$ is the (negative) frequency shift due to the tip-sample interaction.  
The tip-sample distance $z$ was accurately controlled by means of feedback loop regulating the 
$z$ position in such a way that $\Delta f$ was kept constant.
(see Supplementary Material).

Data at the large distance $z=12$ nm, corresponding to $\Delta f=-10$ Hz, show a dissipation peak which is barely visible, corresponding 
to an exceedingly weak van der Waals tip-surface interaction. All other spectra, taken at closer distances, 
exhibit a narrow dissipation peak at a temperature between 114 and 118 K depending upon the surface spot investigated, 
reflecting local changes of $T_c$ determined by inhomogeneous heavy Nb doping,
surface oxygen vacancies, 
and/or stress irregularities. 
The 105K transition temperature of stress free 
pristine
SrTiO$_3$ is notoriously shifted by Nb doping and formation of oxygen vacancies \cite{Ozaki2014}. 
At surfaces, moreover, $T_c$ may under suitable conditions show differences of tens of degrees with respect to the bulk, as seen on SrTiO$_3$(110) \cite{Mishina2000}. 
The dissipation peak in 
this raw data provides a first qualitative confirmation of the suggested connection between 
critical structural fluctuations and AFM dissipation~\cite{Benassi11}.

Fig~\ref{Exp_diss} (b) shows a scanning tunneling microscope (STM) atomically resolved surface 
topography of the 1\% Nb doped SrTiO$_3$(001) surface taken at low temperature $T=5$ K. 
The flat terraces 
are obtained after 
a 2h annealing to $1000^o$ C under ultra 
high vacuum (UHV) \cite{castell02, Silly2006}.% In the antiferrodistorted state, low temperature  SrTiO$_3$ has three types 
%of tetragonal domains, with the tetragonal axis c oriented along [100], along [010], and along [001] directions \cite{lemanov02}. 
Detailed  STM images (see also Supplementary Material) show dark spots (surface defects, perhaps O vacancies \cite{Silly2006}) and  
bright features, decorating what could be edge dislocations \cite{Tanaka93} or other domain walls.

%%%%%%%%%%%%%%%%%%%%%%%%%%%%%%
%%%%%%%%%%%%%%%%%%%%%%%%%%%%%%

We now consider the origin of the pendulum AFM loss process. 
For a start, the tip is sufficiently far from the surface to guarantee that only van der Waals (vdW) (or electrostatic 
if charges were present) tip-substrate interactions are relevant.  Pure SrTiO$_3$ is an insulator and the coupling of a neutral tip must be phononic ~\cite{Volokitin06}. Resistivity measurements of 1\%-Nb doped crystals exhibit %a metallic 
conducting behaviour,
however 
with a carrier density of about $10^{20} cm^{-3}$ \cite{Spinelli2010}, orders of magnitude below that of a good metal. 

\begin{figure}[!ht]\centering
\includegraphics[width=.48\textwidth]{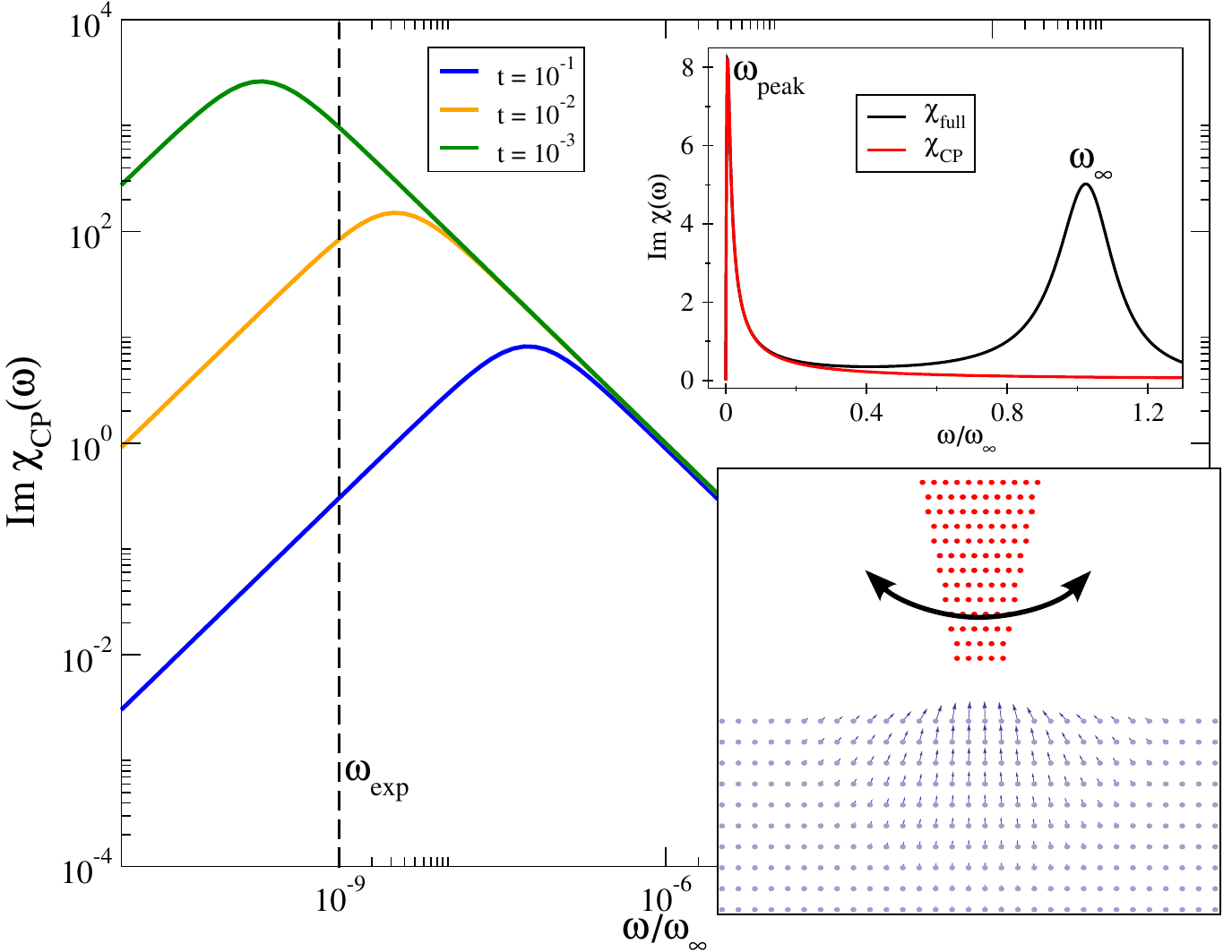}
\caption{\label{chi:fig} $\Ima\chi_{\rm CP}(\omega)$, the low-energy central peak component of
$\Ima\chi(\omega)$ in log-log scale for various temperatures, showing a peak at 
$\omega_{\rm peak}(T)\sim t^{2\gamma}$ that moves towards $0$ as $T\to T_c$. 
(Upper inset) Sketch of the full $\Ima\chi(\omega)$ in linear $\omega$-scale, showing the broad soft-phonon lorentzian
at $\omega_{\infty}$ with the sharp low-energy central peak. For clarity we used here a high
value of $\omega_{\rm low}$ to show both peaks on the same scale. 
(Lower inset) Cross section of 3D simulation of a tip perturbing a semi-infinite crystal through a vdW interaction. The tip (red dots) is shown as a truncated pyramid where  every atom exerts a $-C/r^6$ vdW potential on crystal atoms (blue dots), that are held together by a harmonic potential. 
Arrows (magnified for clarity) represent on a log scale the atom displacements from the relaxed positions.}
%FRANCO WILL PROVIDE NEW FIGURE WHERE omega IS REPLACED BY  omega/omega_infty  OR SOMETHING LIKE THAT}
\end{figure}

Moreover, Auger electron spectroscopy on SrTiO$_3$-(2x2) surface has suggested that the Nb presence is negligible 
in the near-surface region \cite{Lin2011}, so that the low level metallicity due to Nb doping can be considered irrelevant in our experiment.
Fig.~\ref{Exp_diss} (c) shows the maximum dissipation value against tip-sample separation. 
For a spherical tip oscillating above a solid surface 
the dissipation is proportional to $F^2(z)$, where $F(z)$ is the static force resulting from tip-sample interaction. 
The vdW interaction yields a static force $F(z) \propto z^{-2}$, so that the dissipation due to creation of  
phonons in the solid (acoustic phonons in this case, corresponding to the oscillating strain wave under the tip sketched 
in the inset in Fig.~\ref{chi:fig}) should vary as $z^{-4}$ \cite{Volokitin06, Kisiel2011}. The experimental distance dependence, is indeed
best fit by $z^{-p}$ with $p \sim 4.2$ is in excellent agreement with that expectation.

%%%%%%%%%%%%%%%%%%%%%%%%%%%%%%
%%%%%%%%%%%%%%%%%%%%%%%%%%%%%%

We can now directly relate the observed dissipation to the the critical central peak of SrTiO$_3$. 
The noncontact tip vibrating at $f\approx 11$ kHz and at large distance is a very weak perturbation on 
the underlying SrTiO$_3$. Thus, we can make use, rarely appropriate in nanofriction, of linear response theory. 
Moreover since the AFM perturbation affects a sufficiently large portion of SrTiO$_3$,  we  can approximate 
its response by means of the bulk response of the material ~\cite{Shapiro72, Volokitin06}.
The dissipated tip energy per cycle is, in linear response, proportional to the imaginary 
part of the bulk lattice susceptibility $\chi$, in the form
\be\label{Diss}
W(\omega, T) = W_0 + \alpha k_B T \; \Ima\chi(\omega,T) \;,
\ee
where $W_0$ is the dissipation of the free cantilever ($T$-independent in the considered temperature range),
$\chi$ is an appropriate momentum average of the lattice susceptibility $\chi(q,\omega,T)$,
$\alpha$ is a positive, distance-dependent constant
and the temperature factor originates from the term $\hbar\omega n_B(\omega,T)$, with $n_B$ the Bose function, 
in the experimentally relevant regime $\hbar\omega \ll k_B T$.
Using the form by Shapiro {\it et al.} ~\cite{Shapiro72} which accurately describes neutron scattering,  the order parameter 
(zone boundary)  susceptibility can be written as $\chi(q,\omega,T)=\left[\Omega^2(q)-\omega^2+\Pi(q,\omega,T)\right]^{-1}$, 
where $\Omega$ is a bare soft phonon frequency far from the transition and 
$\Pi\sim \Delta(T)-i\omega\Gamma_0(T)$ is a self energy renormalization from anharmonic effects 
(we shall from now on drop the wave-vector $q$ dependence of these quantities).
This simple form of $\Pi$ would lead, in the standard textbook description of a displacive transition~\cite{Kittel:book} 
to a $T$-dependent shift of $\Omega$, resulting in a lorentzian peak in $\Ima\chi (\omega)$ at 
$\omega_{\infty}(T)=\sqrt{\Omega^2+\Delta(T)}$, of width $\Gamma_0$, such that $\omega_{\infty}(T) \to 0$ at $T=T_c$. 
However, the neutron data of SrTiO$_3$ showed that phonon softening is incomplete, $\omega_{\infty}(T_c)\approx 0.5$ meV, 
but accompanied by an extra feature centered at some very-low-energy $\omega_{\rm low}$, the central peak, 
phenomenologically captured \cite{Shapiro72} by an additional contribution to the self-energy $\Pi$
\be
\Pi(\omega,T) = \Delta(T)-i\omega\Gamma_0(T)-\frac{\delta^2(T)}{1-i\omega/\omega_{\rm low}} \;.
\ee
For $\omega \sim \omega_{\infty} \gg  \omega_{\rm low}$ one recovers the usual soft-phonon lorentzian peak 
at $\omega_{\infty}$, but for $\omega \lesssim \omega_{\rm low}$ a second peak appears, well approximated by  
(see upper inset of Fig.~\ref{chi:fig})
\be\label{CPsusc}
\Ima\chi_{\mathrm{CP}}(\omega)=
\frac{\omega_{\rm low}\delta^2(T)}{\omega_\infty^4(T)}
\frac{\omega}{\omega^2+[\omega_{\rm low}\omega_0^2(T)/\omega_{\infty}^2(T)]^2} \;,
\ee
where $\omega_0^2(T)=\omega_\infty^2(T)-\delta^2(T)$ is the quantity that actually vanishes as $T\to T_c$. 
Indeed, the static susceptibility can be shown to be simply related to $\omega_0^2$
\begin{equation}
\chi(0) =\int \frac{d\omega}{\pi}  \frac{\Ima\chi(\omega)}{\omega} = \frac{1}{\omega_0^2(T)} \sim t^{-\gamma} \;.
\end{equation}
The divergence of the order-parameter susceptibility $\chi$ with an exponent $\gamma$, as the reduced temperature 
$t=\md{T-T_c}/T_c$ goes to $0$ is a standard result of the theory of critical phenomena. 
The critical behavior of SrTiO$_3$ is in the 3D-Ising universality class, for which $\gamma\sim 1.24$~\cite{Topler77}.
The low-energy susceptibility $\Ima\chi_{\mathrm{CP}}(\omega)$ of Eq.~\eqref{CPsusc} 
displays a sharp peak at a frequency $\omega_{\rm peak}(T)=\omega_{\rm low}\omega_0^2(T)/\omega_{\infty}^2(T)$,
which moves towards $0$ as $T\to T_c$. 
We can now consider the temperature dependence of the linear response AFM dissipation at the fixed and very low oscillation 
frequency $\omega_{\rm exp} = 2 \pi f$. As $T \to T_c$ from above the dissipation will {\em increase}, 
roughly as $t^{-2\gamma}$, because $\omega_{\rm exp} \ll \omega_{\rm peak}(T)\sim t^{-\gamma}$, to reach
saturation value at $T=\overline{T}$ such that $\omega_{\rm peak}(\overline{T})\approx \omega_{\rm exp}$.
Essentially $\overline{T}$ (here about 1 K above T$_c$) is the temperature below which CP fluctuations average out. Correspondingly, 
below $\overline{T}$ the dissipation levels off as we can essentially take 
$\Ima\chi_{\mathrm{CP}}(\omega_{\rm exp})\approx \omega_{\rm low} \delta^2(T) / [\omega_{\rm exp}\omega_{\infty}^4(T)]$,
which depends very mildly on $T$. 
(experimental values for $\delta^2(T)$ and $\omega_{\infty}(T)$ given by ~\cite{Shapiro72}). We finally obtain an overall
predicted critical form for the AFM dissipation: 
\begin{equation} \label{fig:eqn}
W=W_0+\frac{U}{1+Vt^{2\gamma}}, \
\end{equation} 
where $U$ and $V$ are positive constants. (In the notation of Ref.~\cite{Shapiro72}, 
$Vt^{2\gamma}=\gamma^2\omega_0^4(T)/(\omega_\infty^4(T)\omega_{\rm exp}^2)$ and at low $t$ the relevant dependence on temperature
is given by the $\omega_0^4(T)$ term. )

%%%%%%%%%%%%%%%%%%%%%%%%%%%%%%
\begin{figure}[!ht]\centering
\includegraphics[width=.48\textwidth]{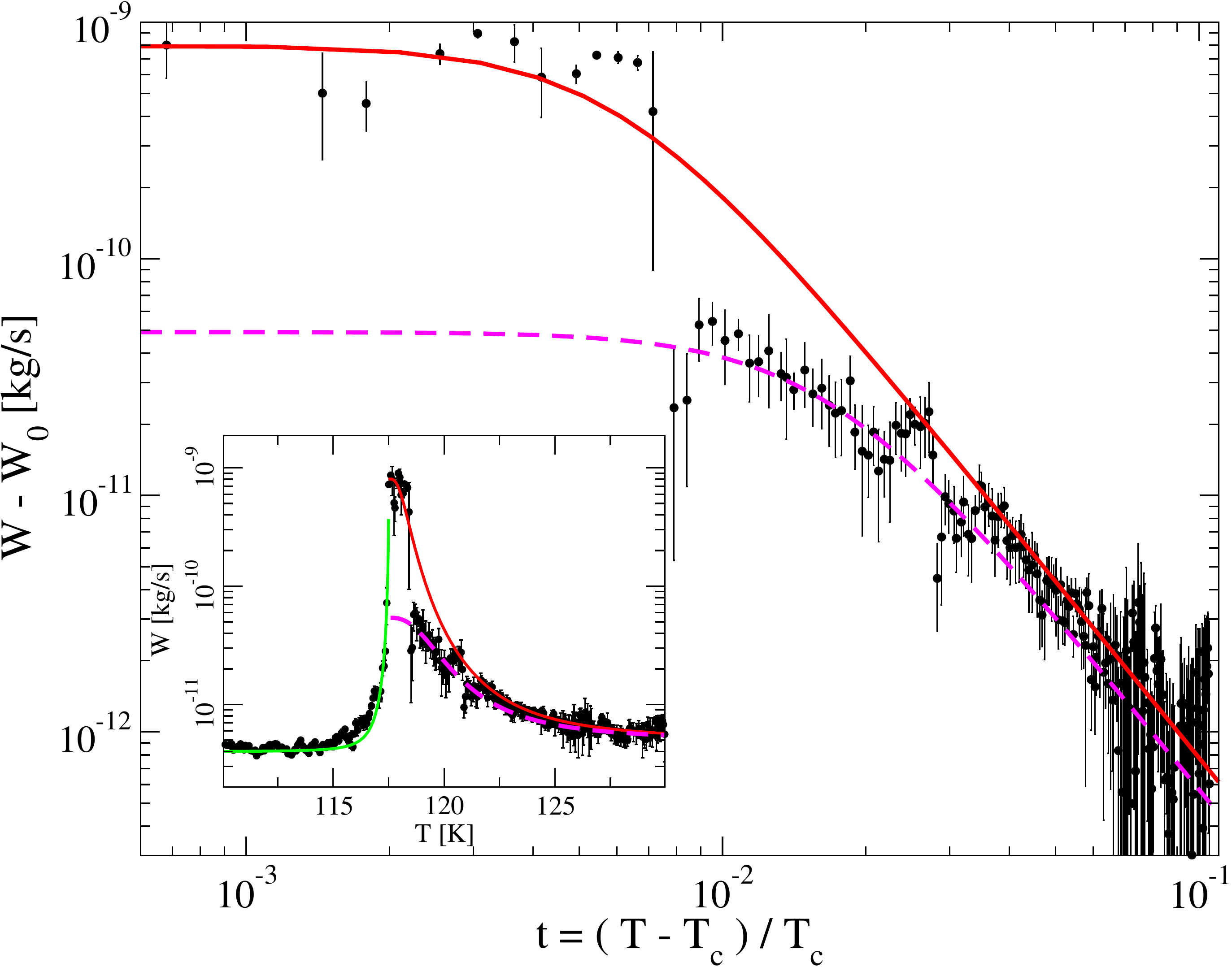}
\caption{\label{W_fit} Experimental dissipation $W$ above $T_c$ as a function of temperature (black dots). 
Inset: the same data in a linear temperature scale, showing data both below and above $T_c$. 
Red and green lines: fit above and below $T_c$ according to Eq.~\eqref{fig:eqn}; pink dashed line: fit above
$T_c$ excluding the high plateau data. The value of $\omega_{\rm low}$ is 156kHz (solid curve) and 44kHz (dashed curve),  
both within a factor 2 of 83 KHz, the value obtained from a simple fit of the saturation temperature (see text). {$\bar{t}$ is roughly 
the temperature where, upon cooling from above $T_c$ the dissipation levels off to a plateau. 
Horizontal error bars corresponding to $\delta t \sim 10^{-4}$ are small and omitted.  }
}
\end{figure}
%%%%%%%%%%%%%%%%%%%%%%%%%%%%%%
%
Fig.~\ref{W_fit} shows on a log-log scale the data for $W-W_0$ at $\Delta f=-40$Hz ($z \sim 3.5$ nm)  
for $T>T_c\approx 117.58$ K. Considering the experimental uncertainty mainly due to noise in the dissipation signal, a slope $t^{-2\gamma}$ provides a 
good fit well above T$_c$, followed by a saturation when $t<\bar{t}\sim 10^{-2}$.
Taking from Ref~\cite{Shapiro72} $\omega_{\infty}^2\approx 0.3$ meV$^2$ and $\omega_0^2(\bar{t})\approx 0.04$ meV$^2$, 
we finally observe that this saturation of AFM dissipation determines the low-energy width parameter $\omega_{\rm low}$ 
as $\omega_{\rm low}=\omega_{\rm exp} \omega_{\infty}^2(\bar{t})/\omega_{0}^2(\bar{t})\sim 83$ kHz. 

We draw in summary four conclusions. First, bulk structural phase transitions are indeed revealed 
by AFM dissipation, as was predicted.~\cite{Benassi11} Strikingly, in the present noncontact realization, this is realized without literally touching the 
crystal. Second, the pendulum AFM dissipation picks up precisely the long debated central peak fluctuations, 
here responsible for the dissipation at the extremely low AFM pendulum frequency of 11 kHz. 
Third, the unknown 
breadth $\omega_{\rm low}$ of the central peak in the dynamical structure factor 
$S(\omega)=\Ima\chi(\omega)/\omega$ 
now  obtained as an intrinsic property of SrTiO$_3$ 
is about 80 kHz, well below the 
upper bound set by the neutron resolution limit of $6$ MHz. 
This CP width is manifested in AFM dissipation as a peak at 
$\omega_{\rm peak}(T)=\omega_{\rm low} \omega_{0}^2(T)/\omega_{\infty}^2(T) \approx 3.2 \; t^{\gamma}$ MHz. 
Fourth, the noncontact, large distance tip-surface coupling elicits a phononic dissipation attributable in turn 
to a slowly varying tip-induced strain, and not to the primary antiferrodistortive 
order parameter, to which the far away tip 
and its motion cannot directly couple. While this realization does not reveal by itself the intimate nature of the CP, 
which remains open to discussion~\cite{Cowley96}  it does show that the exceedingly slow critical CP fluctuations 
must involve a large component of  strain, which is the secondary and not the primary order parameter of the structural transition.

The Basel group acknowledge financial support from the Swiss National Science Foundation (NSF), 
the SINERGIA Project CRSII2 136287/1 and the Swiss National Center of Competence in Research on “Nanoscale Science” (NCCR-NANO).
The SISSA group acknowledges research support by MIUR, through PRIN-2010LLKJBX\_001, 
by SNSF, through SINERGIA Project CRSII2 136287/1, and the ERC Advanced Research Grant
N. 320796  MODPHYSFRICT.

\end{document}